\RequirePackage{fix-cm}

\documentclass[a4paper,10pt]{article}   
\usepackage{graphicx}
\begin{document}
\title{Horizontal rotation signals detected by ``G-Pisa'' ring laser for the $M_w$=9.0, March 2011, 
Japan earthquake}

\author{J.~Belfi$^1$ \and N.~Beverini$^1$ \and G.~Carelli$^1$ \and A.~Di~Virgilio$^2$
\and E.~Maccioni$^1$ \and G.~Saccorotti$^3$ \and F.~Stefani$^1$ \and A.~Velikoseltsev$^4$}





\maketitle
\begin{center}
\it{$^1$ Department of Physics ``Enrico Fermi'', Universit\`a di Pisa, Pisa, Italy\\
$^2$ INFN Section of Pisa, Pisa, Italy\\
$^3$ INGV - Pisa, Italy and Fondazione Prato Ricerche - Prato, Italy\\
$^4$ Department of Laser Measurement and Navigation Systems, St.-Petersburg Electrotechnical University, St. Petersburg, Russia}
\end{center}
\begin{abstract}
We report the observation of the ground rotation induced by the $M_w$=9.0, 11th of March 2011,
Japan earthquake. The rotation measurements have been conducted  with a ring laser gyroscope operating in a vertical plane,
thus detecting rotations around the horizontal axis. 
Comparison of ground rotations with vertical accelerations from a co-located
force--balance accelerometer shows excellent ring laser coupling at periods longer than 100s.
Under the plane wave assumption, we derive a theoretical relationship between horizontal rotation 
and vertical acceleration for Rayleigh waves. Due to the oblique mounting of the gyroscope with respect 
to the wave direction--of--arrival, apparent velocities derived from the acceleration / rotation rate ratio
are expected to be always larger than, or equal to the true wave propagation velocity. This
hypothesis is confirmed through comparison with fundamental--mode, Rayleigh wave phase velocities 
predicted for a standard Earth model. 
\end{abstract}

\section{Introduction}
High-sensitivity ring laser technology \cite{Stedman1997} 
that was developed for geodetic purposes in the past decades (see e.g. 
\cite{Rautenberg1997,Schreiber2003a}) led to the detection
of earthquake-induced rotational ground motions far from seismic sources
\cite{Stedman1995,McLeod1998,Pancha2000,Stedman2006}
opening a new direction in seismic instrumentation. 
Using the C-I ring--laser apparatus in
Christchurch, New Zealand, these authors were the first to exploit the simple 
relationship between transverse acceleration and rotation rate
assuming plane, horizontally--polarized wave propagation. Nevertheless,
their observations were not fully consistent, and were only
studied in a narrow frequency band. After then, detailed analyses of
co-located measurements of transverse acceleration and rotation rate
were carried out by \cite{Igel2005}, \cite{Igel2007} and \cite{Cochard2006}. 
They showed that for several distant large earthquakes the waveforms
and amplitudes of translations and rotations matched the expected
behavior assuming plane wave theory. 
In all these past studies, the experimental setup was arranged in 
order to be sensitive to rotation around the local vertical axis, 
sometimes also called twist, yaw, spin or torsion in engineering 
applications.

Rotations around horizontal axis (tilt) couple in a different way with 
seismic waves. With the exception of a single observation briefly mentioned by \cite{Pancha2000} 
for the G0 ring laser at Cashmere, New Zealand, horizontal rotations have 
no yet been detected systematically using ring-laser systems \footnote{An other ring laser operating with the vertical plane is the  PR-1 ring laser. It 
is mounted on the wall of a laboratory in Department of Physics and Astronomy 
building at University of Christchurch. However its major goal is the detection of seismically induced rotations in buildings, and it's not operating continuously}. 

Since June, 2010, a laser gyroscope with a perimeter of 5.20 m, named ``G-Pisa'', is operating 
at the site of the Virgo gravitational wave interferometer, located in Cascina (Pisa, Italy).
More details about the present status of the ``G-Pisa'' apparatus are presented in another 
paper of this same issue \ref{Belfi_Gpisa}. The high sensitivity and low noise level in the spectral 
range between 1~mHz and 1~Hz makes ``G-Pisa'' an excellent tool for detecting rotational signals associated
with the passage of seismic waves. 

Initially, the optical cavity of ``G-Pisa'' was arranged horizontally. In this configuration, 
it was sensitive to rotations around the vertical axis and  recorded several earthquakes at local 
and regional distances. Starting on December, 2010, the gyroscope layout was turned, bringing the laser 
cavity in a vertical plane, in order to be sensitive to rotation motion around an horizontal axis. 
In this configuration the gyroscope recorded the large $M_w$=9.0 earthquake
which struck the east coast of Honshu, Japan, on 2011 March 11.\\

This paper focuses entirely on the observations of this catastrophic event and aims at
extending the previous studies to the analysis of seismically--induced horizontal ground rotations. 
The paper is structured as follows. First, we describe the experimental apparatus and present the data set.
Then, we use plane--wave theory to derive a theoretical relationship between vertical accelerations and 
horizontal rotations associated with the propagation of Rayleigh waves. Finally, we use 
narrow-band correlation analysis to derive apparent Rayleigh waves phase velocities,  which are compared 
to those predicted for a standard Earth model.

\section{Experimental apparatus}
Large-frame laser gyroscopes are the most sensitive devices among the optical rotation sensors.
These sensors are free from moving parts and in principle measure rotations completely 
rejecting linear accelerations and gravitational force and, when
properly conditioned, provide excellent low-frequency performances. 
They consist essentially of gas ring-lasers operating in continuous-wave and 
single-mode for both the two counter-propagating light beams. When the ring structure 
is rotating with respect to an inertial frame,
the cavity resonance conditions  for the two beams are separated in frequency. 
The optical detection of the rotation rate is performed by  measuring  the beat signal between 
the two counter--propagating beams outside the cavity.
The beat frequency (Sagnac frequency) is related to the rotation rate by a scale factor. 
For a ring laser with perimeter $P$, 
area vector $\vec{A}=A \hat{n}$ and  wavelength $\lambda$, the angular velocity $\Omega$ of its reference frame
induces a Sagnac frequency given by:

\begin{equation}
\label{primaeq}
f_{S}=\frac{4 A}{P \lambda}\vec{\Omega}\cdot{\hat{n}}.
\end{equation}

Very small rotational signals are observed by using the Earth rotation rate  as  bias. For a gyro located at the latitude 
$\phi$, the Earth induced frequency shift for the two counterpropagating optical beams can be expressed as:

\begin{equation}
\label{sagnac_geografico}
f_{S}=\frac{8 \pi A}{P \lambda T_{l.o.d.}}\,\sin{(\phi+\theta_{V})}\,\cos{(\theta_{N})},
\end{equation}

 where  $\theta_{V}$ and $\theta_{N}$ are respectively the angles formed by the 
laser area normal vector with the local Earth radius  and the local meridian while 
$T_{l.o.d.}$ is the time length of the sidereal day.

The rotational signal from the earthquake  has been detected with the ``G-Pisa'' 
gyrolaser, a He-Ne laser  emitting on the red line at $\lambda=\rm{632.8~nm}$ and 
operating in a squared cavity $1.82~\rm{m^2}$  in area. 
At the moment of the earthquake event, 
``G-Pisa'' was oriented  with the area vector horizontally aligned along the 
Virgo north-south arm. 

\begin{figure}
\begin{center}
 \includegraphics[scale=.5]{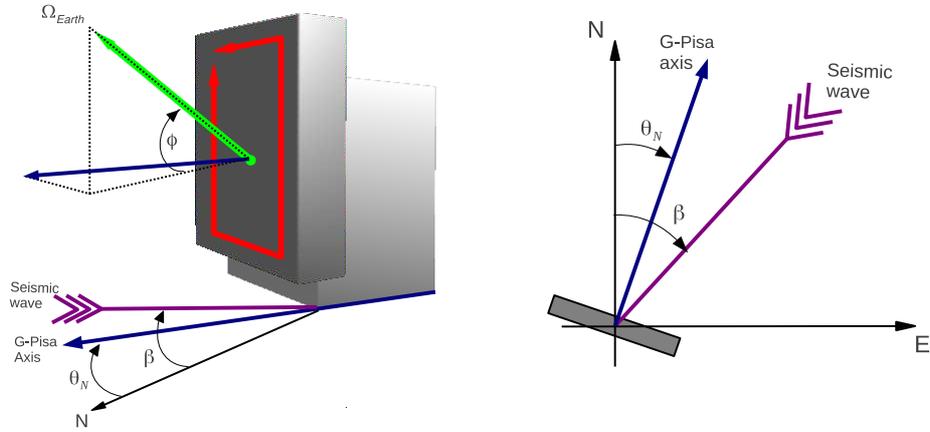}
\end{center}
\caption{The ``G-Pisa'' apparatus in the  vertical plane configuration: Left: schematics of the experimental apparatus. Right: top view. 
The reported angles are defined in the text.}
\label{fig:JEQscheme.eps}
\end{figure}

A scheme of the experimental setup is sketched in figure 
\ref{fig:JEQscheme.eps}. 

From eq.(\ref{sagnac_geografico}), by considering the latitude of Cascina $\phi=43^{\circ}37^{'}53^{''}$, 
one obtains an expected mean value for the Sagnac frequency 
due to the Earth rotation bias, of about 106.2 Hz. The experimentally estimated  value differs for few 
parts per thousand from the expected one. Such a  difference 
has to be  attributed to a slight misalignment between  the laser  plane and the earth radial direction  
as well as to  systematic errors arising from the laser dynamics. \\

Several seismological instruments are currently operating in different sectors of Virgo's
premise. The most sensitive of these is a Guralp CMG40-T broad-band seismometer. Unfortunately,
by the time the earthquake occurred, this instrument was operated  with a high gain 
level in order to monitor the microseismic activity with high sensitivity, 
and thus the earthquake signal resulted severely clipped. To the purpose 
of comparing the translational components of ground motion to the rotational ones, we therefore
make use of signals recorded by a tri--axial Episensor ES-T, force--balance accelerometer located 
at Virgo's detection bench, some 10 m apart from the ring--laser apparatus.\\

The Y and X components of the accelerometer are oriented along Virgo's N arm (see Fig. 
\ref{fig:JEQscheme.eps}) and perpendicularly to it, respectively. 
Given the frequency range (and thus wavelength) considered in this study, the two instruments
are considered as co--located.

\subsection{Data acquisition}
The beat frequency between the two counter propagating gyrolaser beams is detected by a photodetector 
and acquired  by the Virgo data acquisition system at the rate of $5~\rm{kHz}$.
The rotation rate signal is then reconstructed up to the frequency of 50 Hz by using the following  procedure:
\begin{itemize}
\item[-]the optical-beat signal $v(t)$ is bandpass-filtered (with a digital first order Chebyshev filter) around 
the nominal Sagnac frequency within a bandwidth of about 90 Hz;
\item[-]the filtered signal and its Hilbert transform $w(t)$ are combined to form the analytic signal $z(t)=v(t)+j w(t)$;
\item[-]by expressing  the analytic signal in the form $z(t)=|z(t)|e^{j \psi(t)}$, 
the instantaneous phase is estimated as $\psi(t)\equiv 2 \pi f_S t$  ;
\item[-]the instantaneous frequency $f_S$, connected to the 
rotation rate by the eq.(\ref{primaeq}) is finally estimated by calculating the derivative of the unwrapped instantaneous phase
$\psi$;
\item[-]the instantaneous frequency is down-sampled down to the final sampling frequency of $50~\rm{Hz}$ and converted 
in rad/s by using eq.(\ref{sagnac_geografico}).
\end{itemize}

Signals from the accelerometer are acquired by Virgo's internal acquisition system, and successively 
down--sampled to 50 Hz after appropriate low--pass filtering.
  
\section{Data analysis}
Data from ``G-Pisa'' represent the rotation rate $\Omega_\mathbf{x}$ (tilt) around a 
horizontal vector $\mathbf{x}$ which is oriented $N19^{\circ}E$ (Fig. \ref{fig:JEQscheme.eps}). 
The expected backazimuth (receiver-to-source angle measured clockwise from the N direction) of the Japan earthquake seismic wave is about 36$^\circ$. 
Thus,  the area vector of ``G-Pisa'' is misoriented by
about 73$^\circ$ with respect to the rotation vector of the Rayleigh waves (Fig. \ref{fig:JEQscheme.eps}).
This oblique mounting implies that the recorded rotation rates can be significantly smaller
than those predicted by theory.\\

\begin{center}
\begin{figure}
 \includegraphics[width=12cm]{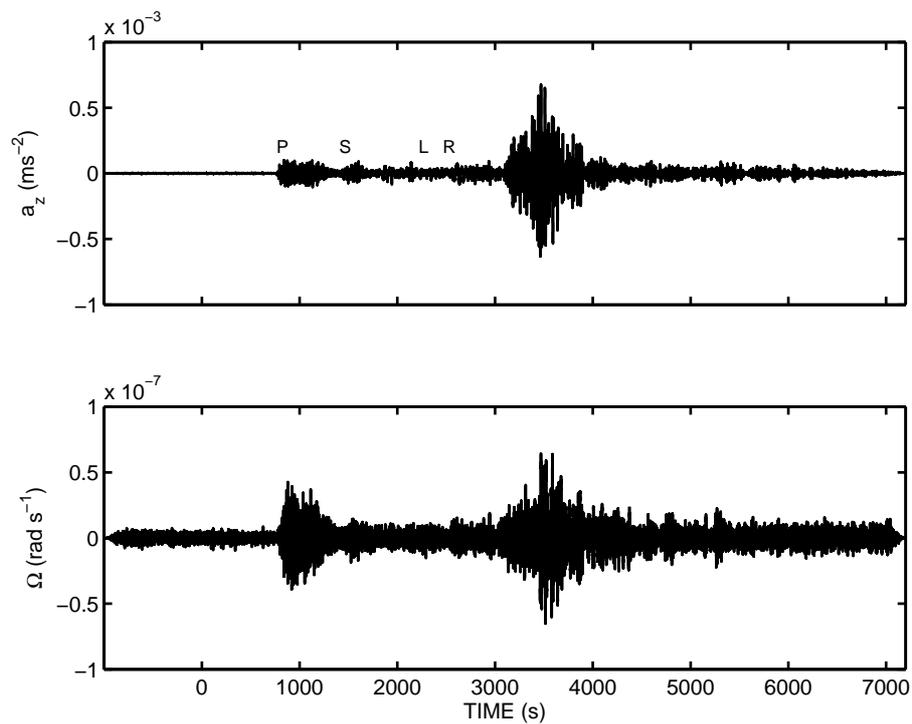}
\caption{Recordings of vertical--component ground acceleration (top) and rotation rate (bottom) of 
the March 11, 2011, $M_w$=9 Sendai-Honshu earthquake. Instruments are located within few meters  inside
the Virgo gravitational observatory laboratories. Traces have been band-pass filtered over the 100--s
1--s period range using a 2-pole, 2-passes
Butterworth filter. The main seismic phases are marked; P,S,L and R correspond to direct P, S, 
Love and Rayleigh waves, respectively. Time 0 corresponds to the origin time of the earthquake
(05:46:23 UTC).}
\label{fig:Fig_timeseries}
\end{figure} 
\end{center}

Figure \ref{fig:Fig_timeseries} shows the simultaneous recording of the vertical component of ground acceleration $\ddot{u}_z$
and the horizontal rotation rate $\Omega_x$ from ``G-Pisa''. Overall, waveform morphology of the two time series are significantly similar. 
The main arrivals (P-, S- and Rayleigh waves) are clearly detectable on both recordings.\\

\subsection{Spectral Properties}

\begin{center}
\begin{figure}
\includegraphics[width=12cm]{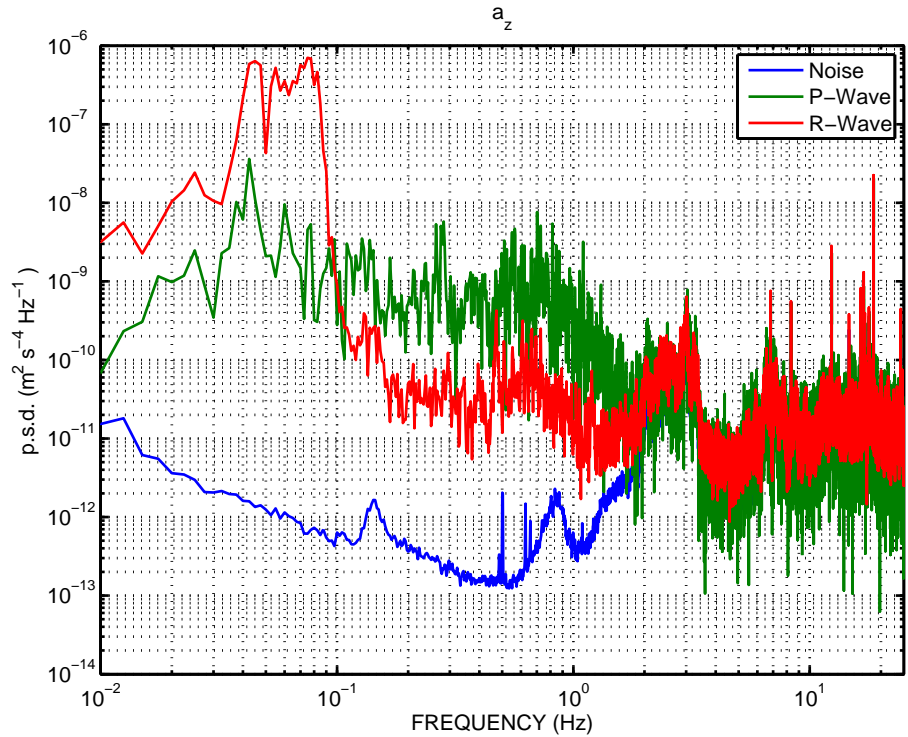}
\caption{Power spectral estimates for the $a(t)$ time series of a 5-hour-long noise window preceding
the earthquake, a 600-s-long window encompassing the P-wave arrivals and coda, and a 1800-s-long window
starting at the Rayleigh wave arrival. Power spectra are obtained by averaging and then squaring
individual spectral estimates obtained over a 100-s-long window sliding by 50\% of its length along 
the selected signal segments. Before transformation, individual signal slices were detrended, 
demeaned, and tapered by a 5\% Tukey window.}
\label{fig:Fig_NPR_ACC}
\end{figure} 
\end{center}

\begin{center}
\begin{figure}
\includegraphics[width=12cm]{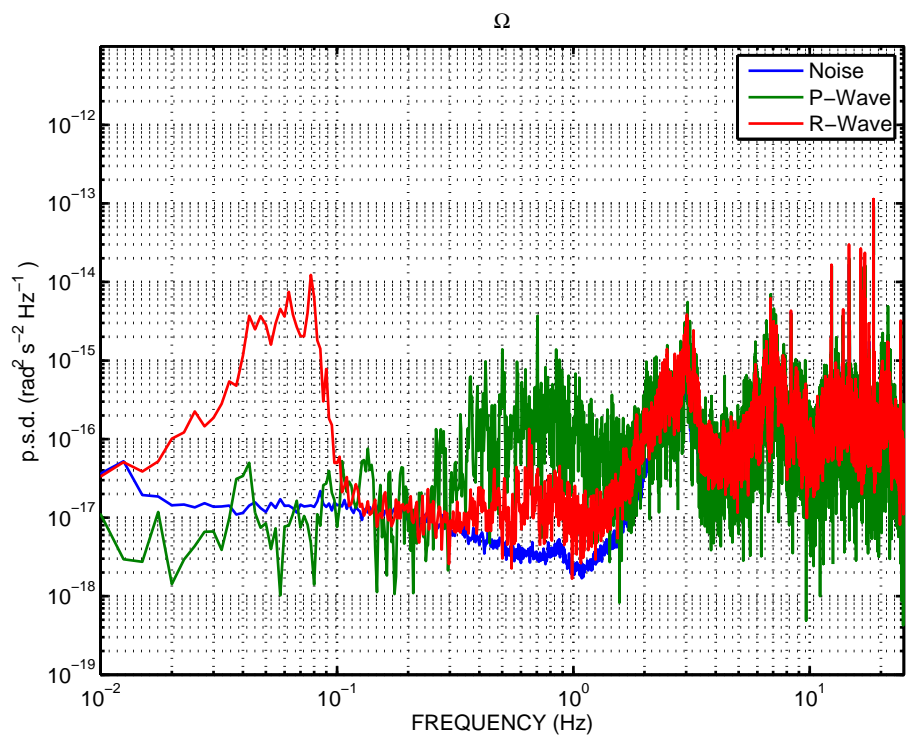}
\caption{The same as in Figure \ref{fig:Fig_NPR_ACC}, but or the Gyrolaser signal.}
\label{fig:Fig_NPR_GYR}
\end{figure} 
\end{center}

Figures \ref{fig:Fig_NPR_ACC} and \ref{fig:Fig_NPR_GYR} illustrate the power spectral densities 
for the background noise preceding the earthquake, the P wave-train and its coda, and Rayleigh waves
as recorded at the accelerometer and gyroscope, respectively. 

Acceleration spectral peaks from both P- and Rayleigh waves span 
the 0.02 Hz - 3 Hz frequency band, and their energy is up to 4-5 orders of 
magnitude larger than that associated with background noise. Once compared to the noise spectrum, the 
rotational earthquake signal is much weaker. P-wave energy is at maximum 2 orders of
magnitude larger than noise power over the 0.2-2 Hz frequency interval; Rayleigh wave
power is 2.5 orders of magnitude and less than one order of magnitude above the background level 
for the 0.03-0.1 Hz and 0.3-2 Hz frequency bands, respectively. \\

The frequency peak between 2 and 4 Hz is a persistent feature at both instruments. 
During the earthquake, its amplitude is slightly larger than during the preceding noise. 
This observation can be interpreted in terms of a site effect and/or resonance of the 
building, inducing narrow-band amplification of the earthquake signal. 
Alternatively, it could be that the microseismic noise 
(mostly of human origin) is stronger during the earthquake, simply because of the 
daily working cycle.\\

\subsection{Relationships Between Ground Translations and Rotations}

For a seismic wavefield defined by the displacement vector $\mathbf{u}(u_{x},u_{y},u_{z})$, 
the relation between infinitesimal rotations $\theta$ and translational motion is obtained through 
application of the curl operator as:

\begin{equation}
\left(
\begin{array}{c}
\theta_x \\
\theta_y \\
\theta_z 
\end{array}
\right)
= \frac{1}{2} \nabla \times {\mathbf{u}} =
\frac{1}{2} 
\left(
\begin{array}{c}
\frac{\partial{{u}_z}}{\partial{y}} -\frac{\partial{{u}_y}}{\partial{z}} \\
\frac{\partial{{u}_x}}{\partial{z}} -\frac{\partial{{u}_z}}{\partial{x}} \\
\frac{\partial{{u}_y}}{\partial{x}} -\frac{\partial{{u}_x}}{\partial{y}} 
\end{array}
\right)
\label{eq:eq01}
\end{equation}

Most of the experiments on the direct ring laser measurement of rotational ground motion 
are based upon sensors measuring rotation rates around a vertical axes (i.e., $\dot\theta_z=\Omega_z$). 
These devices are thus  sensitive to transverse waves polarised on the horizontal (x,y) plane, 
that is S- and Love-waves.
Under these conditions, \cite{Stedman1995} showed that the transverse ground acceleration $\ddot{u}_T$ 
for Love waves is related to the vertical rotation rate $\Omega_z$ by the simple relationship:

\begin{equation}
 \ddot{u}_T = 2 c_{L} \Omega_z,
\label{eq:eq02}
\end{equation}

where $c_{L}$ is the local Love phase velocity. A similar expression is derived for 
the case of S-waves, with an additional factor accounting for the effects of non-horizontal
incidence \cite{Li2001}. The above 
relationship has been exploited by several authors \cite{Igel2005,Igel2007,Takeo2009} 
to derive estimates of the medium velocity from Love waves dispersion characteristics. \\

By the same token, rotations around one of the two horizontal axes (let's say, the $x$-axis) 
requires wavetypes propagating along and transversely-polarised on the vertical ($y,z$) plane.
These are (i) P-wave, as a consequence of wave reflection-conversion at the free-surface; (ii) SV waves, and
(iii) Rayleigh waves.\\

To the purpose of investigating the frequency-dependent performance of the
rotational sensor, we focus on the ground rotation and vertical-component acceleration 
associated with the Rayleigh wave packet. We select these particular wave type and ground motion component
as they are the ones exhibiting the largest amplitude, and hence the best signal-to-noise ratio. In the following,
we thus proceed deriving a theoretical relationship between Rayleigh wave vertical accelerations and rotations.\\ 

Assuming the surface corresponds to the xy plane, the zero traction boundary condition at 
the free surface implies that $\sigma_{iz}~~(i = x, y, z)=0$. Direct
application of Hooke's law in a homogeneous, isotropic medium thus leads to:

\begin{equation}
\frac{\partial{{u}_y}}{\partial{z}} =- \frac{\partial{{u}_z}}{\partial{y}}
\label{eq:eq02a}
\end{equation}

The Rayleigh--wave vertical displacement in a simple half-space Poisson solid is \cite{Lay1995}:

\begin{equation}
{u}_{z}= f(y-c_{R} t) 
\label{eq:eq03}
\end{equation} 

where $c_{R}$ is the Rayleigh wave phase velocity. 
From equations (\ref{eq:eq01}) and (\ref{eq:eq02a}), the horizontal rotation rate is:

\begin{equation}
\Omega_x =\frac{\partial \dot{u}_z}{\partial y}=-\frac{1}{ c_R} \ddot{u}_z
\label{eq:eq05}
\end{equation} 

This latter relationship indicates that the amplitude ratio between the  horizontal rotation rate and the  vertical acceleration,
is equal to the Rayleigh--wave phase velocity $c_R$.
Note that this expression is slightly different from that reported in \cite{Stedman1995},
who did not provide separate expressions for the two  components of ground acceleration.\\
Eq. \ref{eq:eq05} is  a simplification, valid for an infinite half-space. For the more realistic case of a layered Earth, one should  account for the 
dispersion effects which imply a frequency dependence of $c_{R}$. In addition, interference of multiple, higher-order propagation modes 
should also be taken into account \cite{kurrle}.

\subsection{Correlation of Rotation Rates and Vertical Accelerations}

\begin{center}
\begin{figure}
\includegraphics[width=12cm]{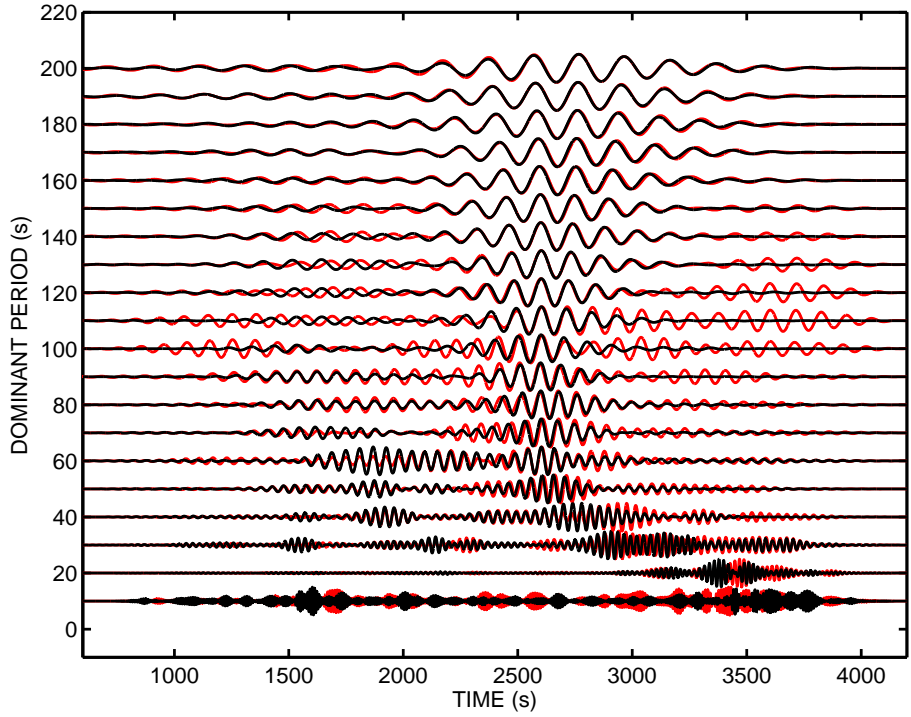}
\caption{Superposition of vertical acceleration (black) and rotation rate (red) 
as a function of dominant period after narrow-band filtering. Each trace has been 
individually normalized. Signals start 1800s before the expected Rayleigh wave arrival; 
the time scale is referred to the event's origin time.}
\label{fig:MFT_Traces}
\end{figure} 
\end{center}

In order to investigate the performance of the
rotational sensor compared with the accelerometer, we
superimpose vertical acceleration and rotation rate after filtering
with a narrow bandpass (2-poles, zero-phase Butterworth with corner frequencies 
0.9 $\times$ 1/T Hz and 1.1 $\times$ 1/T Hz, where T is the period
in seconds). The selected signal segments are 1-hour-long, and they
are centered  at the predicted Rayleigh wave arrival ($\approx$ 2400s after origin time).
The procedure is iterated for 20 center periods spanning the 200s-10s 
period range. At periods longer than 100s, the results indicate excellent 
matching between ground rotation and acceleration (Fig. \ref{fig:MFT_Traces}). 
At shorter periods (20s-100s), the two time series exhibit occasional phase
differences, likely due to interference with late-arriving, scattered  body waves. Nonetheless, the general waveform morphology is still significantly 
similar and the typical dispersive behavior of surface waves is clearly observed
on both recordings.\\ 

\begin{center}
\begin{figure}
\includegraphics[width=12cm]{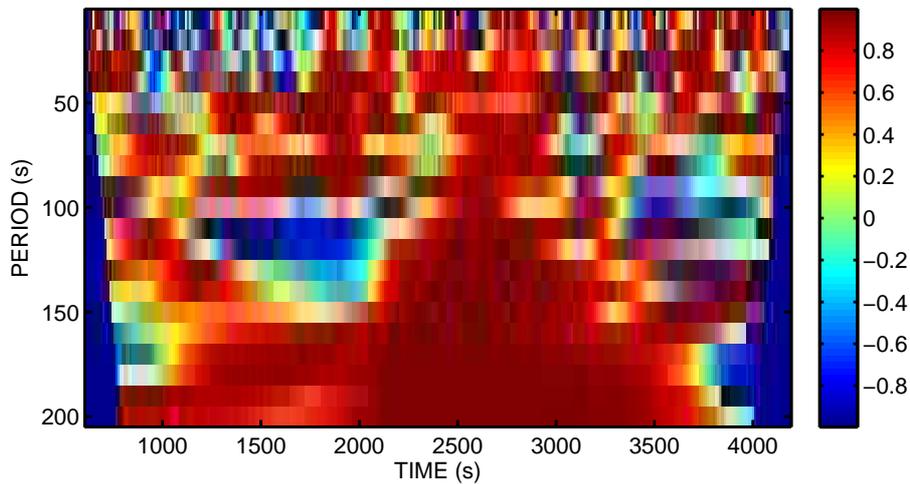}
\caption{Normalized zero-lag correlation coefficient between vertical acceleration and rotation
for the traces shown in Figure \ref{fig:MFT_Traces}. For each frequency band, correlation has been computed over
windows of length 2 dominant periods, sliding along the traces with 50\% overlap.}
\label{fig:MFT_XCorr}
\end{figure} 
\end{center}

For each frequency band, we quantify the time- and frequency-dependent
similarity between rotation rate and vertical acceleration by sliding a time window 
of length twice the dominant
period along the time-series, and calculate the zero--lag normalized
correlation coefficient that is defined between -1 and 1. 
Results are reported in the scaled-color image of figure \ref{fig:MFT_XCorr}, where  
the correlation coefficients are shown in their dependence on both time and periods.

From this latter data matrix, we then select the time intervals and frequency bands 
where the correlation coefficients are greater than an arbitrary 
threshold of 0.95. For those intervals, we take the median of the instantaneous (i.e., sample-by-sample) ratios 
between acceleration and rotation rate as an estimate of the amplitude relationships 
between the two quantities.\\

In general, the oblique mounting of ``G-Pisa'' with respect to the backazimuth implies that
the measured rotations can be significantly smaller than those predicted by theory. Therefore,
if the analyzed portion of the wavefield were purely composed by Rayleigh waves, then
the acceleration / rotation amplitude ratio (eq. (\ref{eq:eq06})) should always
be greater than or equal to the local Rayleigh phase velocity. We thus use the AK135
Earth Model \cite{Kennett1995} to calculate the 
fundamental-mode, dispersion function for Rayleigh-wave phase velocity, and compare this function
to the apparent velocities derived from the $a_z / \Omega_x$ ratio.\\

\begin{center}
\begin{figure}
\includegraphics[width=12cm]{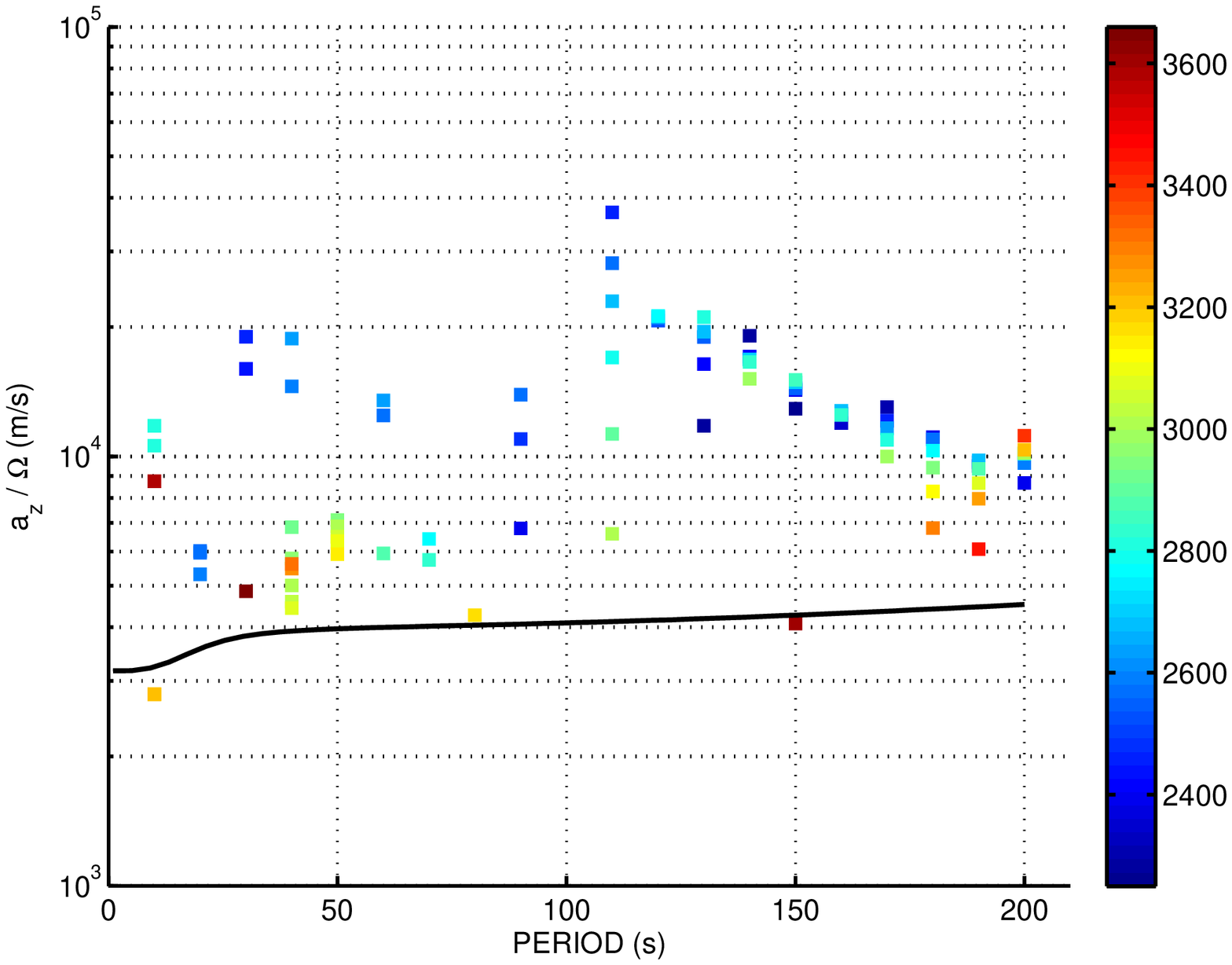}
\caption{Amplitude ratios (apparent phase velocities) between vertical acceleration and rotation 
rate as a function of dominant period, for signal windows depicting zero-lag correlation greater than 0.95. 
The color of the symbols indicate the time location of the analyzed signal window, according to the color 
scale at the right.
The black line marks the Rayleigh-wave phase velocity dispersion curve calculated for the 
AK135 velocity model.  
This line represents a lower bound on
the possible $a_{z} / \Omega$ ratios measured for waves propagating along different directions.}
\label{fig:MFT_AmpRatio}
\end{figure} 
\end{center}

The results, displayed in figure \ref{fig:MFT_AmpRatio} indicate that the 
apparent velocities derived from the acceleration / rotation rate ratios 
are always greater than those predicted by the theory, thus confirming 
the interpretation of the ring--laser signal in terms of ground tilt
associated with the passage of teleseismic, long--period Rayleigh waves.

\section{Discussion and Conclusions}

In this study, for the first time it is shown a remarkable consistency
between rotational ground motions around a horizontal axis and vertical acceleration 
primarily quantified by the cross-correlation coefficient that is calculated in sliding 
time windows along the narrow--band filtered time series. 
Over the 100--200~s period range, the ring laser correlation 
with ground acceleration extends throughout the duration of the Rayleigh wave--packet. At 
shorter periods, the correlation between the two time series is much more discontinuous,
pointing to a scattered wavefield constituted by the interference of multiple wave--types.   

Apparent phase velocities estimated by taking the ratio of vertical acceleration and
rotation rate, are compatible with the expected velocities assuming
plane  wave propagation, confirming recent results for vertically--oriented
ring--laser gyroscopes \cite{Igel2005,Igel2007,Cochard2006}. 
This confirms that multi-component observations allow
the estimate of wavefield properties (e.g. phase velocities, 
propagation direction) that otherwise can be achieved with high accuracy
only through multichannel measurements.
Our results open new perspectives for a wide range of applications
for which high--sensitivity, very broad--band measurements of ground tilt are required.
These include the surveillance of volcanic and geothermal areas, the monitoring of 
carbon dioxide capture and storage in geological structures, the control of oil and gas
production plants.


\section{Acknowledgements}
We acknowledge F. Bosi, A. Gebauer, R. Hurst and U. Schreiber for the useful discussions.
A. Velikoseltsev acknowledges the partial support from the Federal Targeted 
Program ”Scientiﬁc and scientiﬁc-pedagogical personnel of 
the innovative Russia in 2009-2013” of the Ministry of Education and Science of the Russian Federation.
%

\end{document}